\documentclass[a4paper]{article}

\usepackage{amsmath}
\begin{document}

\title{\bf  $SO(10)$ model of standard and dark matter}
\author{V.\,V.\,Khruschov$^{1,2)}$\/\thanks{khruschov\_vv@nrcki.ru} \\ 
$^1${\it NRC Kurchatov Institute, 123182
Moscow, Russia}  \\ 
$^2${\it CGFM VNIIMS, 119361 Moscow, Russia}}
\date{}
\maketitle

We consider a novel model for three standard families of left chiral states of quarks and leptons conjointly with a new family of dark matter fermionic particles and a sterile neutrino. It is suggested to use a SO(10) symmetry for description of these families for the first time. It is presented estimations of masses of dark matter particles and the sterile neutrino as well limitations for values of mixing parameters between the new family particles and active neutrinos. This model can be used for predictions and interpretations of results of experiments for sterile neutrino and dark matter particles search.
 


\section{Introduction}
\markboth{\thesection\hspace{1em} Introduction}{\hspace{2em}\it V.V. Khruschov \hspace{2em}SO(10) model of standard and dark matter}
\label{Section1}
At present the Standard Model (SM) of strong, electromagnetic and weak interactions with the $SU(3)_c\times SU(2)_L\times U(1)_Y$ gauge symmetry is the theory of physics for known quarks and leptons. With respect to the $SU(2)_L\times U(1)_Y$ subgroup the quarks and leptons are distributed among left and right chiral multiplets, at that  the left and right multiplets for these particles differ as with dimensions, as with values of the hypercharge $Y$. However it is clear now that SM is the low energy limit of a more general theory, which is named as the Grand Unification Theory (GUT). A GUT should unify three families of quarks and leptons and diminish the number of SM parameters. It is existed different GUT variants, but in most cases they deal with the known standard matter. Recently it is appeared experimental data which cannot be explained in the SM framework. In first turn among them it is data concerning dark matter, neutrino masses, as well as possible existence of sterile neutrinos. 

In the present paper a novel model of standard and dark matter particles with a $SO(10)$ symmetry is considered. The $SO(10)$ symmetry was proposed for the first time in Refs. \cite{Fritzsch,Georgi} as the gauge symmetry of a GUT for standard matter particles belonged with a family. The theory with the $SO(10)$ symmetry is considered by some authors as the basis for the final GUT of standard matter (e.g., see \cite{Minkowski,Bajc,Khruschov1}). Below this symmetry is used in another manner, namely as a symmetry of new interactions among particles of standard and dark matter. We consider conjointly the known families of quarks and leptons and a new family consisted of dark matter fermionic particles and sterile neutrino. One of the main problems of this model is a suppression of transitions among standard matter particles and dark matter particles. In the present paper this problem can be resolved by known means, namely with increasing to a required degree of mass values of gauge bosons being carried out such transitions.
 
The paper is organized as follows: in Section 2 we provide an outline of the novel model with a $SO(10)$ symmetry for particles of standard and dark matter including a sterile neutrino. It is suggested a chain of breakings of the SO(10) symmetry up to a  low energy region. In Section 3 we introduce mixing parameters among active neutrinos, sterile neutrino and dark matter particles. We find estimations of mass values for new particles and limitations for parameters of mixing with new particles. We then conclude with a summary of the results in Section 4. 
\section{Families of standard and dark matter particles and a SO(10) symmetry }
\label{Section2}
Let us consider three families of known quarks and leptons and a new family of dark matter particles including a sterile neutrino. We regard only left chiral states of these particles, which present in a form of 
$(4\times4)$-matrix.
\begin{equation}
C_{L}=\left(\begin{array}{cccc}
(\nu_e)_L & (\nu_{\mu})_L & (\nu_{\tau})_L & (\nu_s)_L\\
(e^-)_L & (\mu^-)_L & (\tau^-)_L & ({\sigma})_L\\
(u)_L & (c)_L & (t)_L & (N)_L\\
(d')_L & (s')_L & (b')_L & (S)_L\end{array}\right),
\label{eq1}
\end{equation}
\noindent where $(\nu_s)_L$ is a left chiral state of a sterile neutrino,  $(\sigma)_L$ is a left chiral state of the  dark matter particle $\sigma$, $(N)_L$ is a left chiral state of the  dark matter particle $N$,  $(S)_L$ is a left chiral state of the  dark matter particle $S$ and so on. 
$(d')_L$, $(s')_L$, $(b')_L$ are left chiral states of quarks $d$, $s$ and $b$ mixed with the CKM matrix. 

We assume that the states (\ref{eq1}) collected into a multiplet $C_L$ realize a 16-dimensional representation of a $SO(10)$ symmetry. This symmetry differs from the $SO(10)$ symmetry suggested in Refs. \cite{Fritzsch,Georgi} and is the group of a new gauge symmetry for interactions of standard and dark matter particles. Let us denote this group as $SO(10)_X$. The particles belonging to the multiplet (\ref{eq1}) have new charges with respect to the $SO(10)_X$ symmetry. These charges do not connected with known electrical charges of the particles (\ref{eq1}). However taking into account the electrical charges we can draw definite conclusions about possible mixing among the particles (\ref{eq1}), this will be done in Section \ref{Section3}. 

$SO(10)_X$ undergoes at several breaking stages in a transition region from super high energy to low energy, where SM holds. We propose the following chain of $SO(10)_X$  symmetry breakings taking into account the suggested above composition of the 16-dimensional $SO(10)_X$ representation $C_L$. 
\begin{eqnarray*}
SO(10)_X \to SU(4)\times SU(2)\times SU(2) \\
\to SU(3)\times U(1)\times U(1)\times U(1) \\
\to SU(2)\times U(1)\times U(1)\times U(1)\times U(1)\\
\to U(1)\times U(1)\times U(1)\times U(1) \times U(1).      
\end{eqnarray*}
So the $SO(10)_X$  symmetry reduces to the $[U(1)]^5$ symmetry, which also breaks at transition in the SM region. 
\section{Estimations of masses and parameters of mixing with active neutrinos for dark matter particles and a sterile neutrino}
\label{Section3}
The great reduction of the $SO(10)_X$  symmetry in a low energy region makes difficult search of $SO(10)_X$ effects in this region. Nevertheless, we can find some estimations of characteristics of dark matter particles $\sigma$, $N$, $S$ and sterile neutrino $\nu_s$ entering into $C_L$ on the base of the reduction chain  of $SO(10)_X$  and existed experimental data \cite{Olive}. First of all, it should be expected, when taking into account large difference between quarks masses and leptons masses, that masses of doublets 
\begin{equation}
x = \left(\begin{array}{c}
\nu_s\\ \sigma \end{array}\right), \quad
X =
\left(\begin{array}{c}
N\\ S\end{array}\right)
\label{eq3}
\end{equation}
\noindent differ in a great extent. So let us regard $\nu_s$ and $\sigma$ as light dark matter particles, while $N$ and $S$  as heavy dark matter particles. We introduce left chiral mass states of neutrino and dark matter fermionic particles by analogy with description of neutrino mixing: 
\begin{equation}
(\nu_1)_L, (\nu_2)_L,(\nu_3)_L,(\nu_4)_L,(\delta_1)_L,
(\delta_2)_L, (\delta_3)_L
\label{eq4}
\end{equation}
\noindent  Mixing states $(\nu_e)_L$, $(\nu_{\mu})_L$, $(\nu_{\tau})_L$, $(\nu_s)_L$, $(\sigma)_L$, $(N)_L$, $(S)_L$ can be written through the mass states (\ref{eq4}). We give seven approximate expressions neglecting evidently small terms. 
\begin{eqnarray*}
(\nu_e)_L = a_1(\nu_1)_L+a_2(\nu_2)_L+a_3(\nu_3)_L+a_4(\nu_4)_L+a'_1(\delta_1)_L,\\
(\nu_{\mu})_L = b_1(\nu_1)_L+b_2(\nu_2)_L+b_3(\nu_3)_L+b_4(\nu_4)_L+b'_1(\delta_1)_L,\\
(\nu_{\tau})_L = c_1(\nu_1)_L+c_2(\nu_2)_L+c_3(\nu_3)_L+c_4(\nu_4)_L+c'_1(\delta_1)_L,\\
(\nu_s)_L = d_1(\nu_1)_L+d_2(\nu_2)_L+d_3(\nu_3)_L+d_4(\nu_4)_L+d'_1(\delta_1)_L,\\
(\sigma)_L = e_1(\nu_1)_L+e_2(\nu_2)_L+e_3(\nu_3)_L+e_4(\nu_4)_L+e'_1(\delta_1)_L,
\end{eqnarray*}
\begin{equation}
(N)_L = f'_2(\delta_2)_L+f'_3(\delta_3)_L,\nonumber
\end{equation}
\begin{equation}
(S)_L = g'_2(\delta_2)_L+g'_3(\delta_3)_L \nonumber
\end{equation}
\noindent Certainly, inequalities written below must be held in order to adjust the expressions for mixed states with existed experimental data for active neutrinos oscillations \cite{Olive, Capozzi, Forero, Gonzalez}: 
\begin{equation}
a_1, a_2, a_3 \gg a_4 \gg a'_1, \nonumber
\end{equation}
\begin{equation}
b_1, b_2, b_3 \gg b_4 \gg b'_1, \nonumber
\end{equation}
\begin{equation}
c_1, c_2, c_3 \gg c_4 \gg c'_1, \nonumber
\end{equation}
\begin{equation}
d_4 \gg d'_1 \gg d_1, d_2, d_3, \nonumber
\end{equation}
\begin{equation}
e'_1 \gg e_4 \gg e_1, e_2, e_3, \nonumber
\end{equation}
\begin{equation}
f'_2 \gg f'_3, \nonumber
\end{equation}
\begin{equation}
g'_3 \gg g'_2. \nonumber
\end{equation}
\noindent The $(7\times 7)$-matrix $U'$, organized with coefficients  $a_i$, $b_i$, $c_i$,
$d_i$, $e_i$, $a'_i$, $b'_i$, $c'_i$, $d'_i$, $e'_i$, $f'_i$, $g'_i$, will be unitary in the considered model. It is the generalization of the known active neutrinos matrix  $U_{PMNS}$ for neutrinos and dark matter particles. One can see that the sterile neutrino 
$\nu_s$ may be thought of as the dark matter fermionic particle, and in turn the dark matter fermionic particles $\sigma$, $N$ and $S$  may considered as additional sterile neutrinos.  

The basic propositions of the $SO(10)$ model for standard and dark matter particles have been presented above. In what follows we give estimations for mass values of the new particles which have their bases in some observational data concerning sterile neutrinos and dark matter particles  \cite{Abazajian, Bulbul, Boyarsky,Drewes, Demianski}. It is significant that specific mass values are chosen much as an illustration for possible application of the model. Variations in these values are of little consequence for the structure and meaning of the considered model. 

Let us adopt the mass value of the particle $\nu_4$  of the order of 1 eV and the mass value of the particle $\delta_1$ of the order of 10 keV. Then it becomes probable the occurrence of shot baseline anomalies of neutrino data \cite{Abazajian} and the detection of 3.55 keV line in gamma-spectra of some astrophysical sources \cite{Bulbul, Boyarsky}. 
 We take the mass values of the particle $\delta_2$ and $\delta_3$  of the order of 1 GeV accepting estimations for heavy right neutrinos from the work \cite{Drewes}.  
 Then a possibility appears to explain the cosmological origin of the baryonic matter in the universe, as well as a number of the observed dark matter (DM) facts \cite{Drewes, Demianski}. However the fermionic dark matter with such mass values of the particles have to be non-stable on account of the restriction obtained in the work \cite{Lee}, or constitute only a small part of dark matter, which will in the main contain bosons in this case \cite{Boehm, Harigaya}.
 
If one consider the suggested model as a model with additional sterile neutrinos \cite{Kopp, Gariazzo, Khruschov2}, then it can be classified as a (3+1+1+2) model. The inclusion of sterile neutrinos with different mass values as in the (3+1+1+2) model can lead to explanation of some complicated problems of supernova physics (see, e.g. \cite{Warren, Yudin}).
\section{Conclusions}
\label{Section4}
The novel chiral model with the gauge $SO(10)$ symmetry for standard and dark matter particles including a sterile neutrino has been considered in the paper. The model predicts existence of  new fermionic particles, namely the doublet of light  particles and the doublet of heavy  particles. The first doublet can consist of a sterile neutrino with a mass of the order of 1 eV and  a dark matter particle with a mass of the order of 10 keV or tens MeV. The second one can consist of two dark matter heavy particles with masses of the order of 1 GeV or higher. The model predicts existence of forty five new gauge bosons, which acquire great masses after $SO(10)$ breaking. It is introduced mixing parameters among active neutrinos, sterile neutrino and dark matter particles  and limitations for parameters of neutrino mixing with new particles are obtained. It is suggested the basic statements of the model, which will be working out later on for prediction and interpretation of data of experimental searches for sterile neutrinos and dark matter particles. 

The author would like to thank  A.G. Doroshkevich, Yu.S. Lutostansky, V.I. Lyashuk, D.K. Nadyozhin, S.V. Semenov, S.V. Fomichev and A.V. Yudin for useful discussions. This work was supported in part by the RFBR grant 14-22-03040 with the code "ofi-m".

\end{document}